\documentclass[11pt]{article}
\usepackage{hyperref}
\pdfoutput=1
\begin{document}
\title{Droplets bouncing over a vibrating fluid layer}
\author{Pablo Cabrera-Garcia$^1$ and Roberto Zenit$^2$ \\
$^1$ Facultad de Ciencias\\
$^2$ Instituto de Investigaciones en Materiales \\ Universidad Nacional Aut\'onoma de M\'exico \\
Cd. Universitaria, M\'exico D.F., 04510 \\ M\'EXICO} \maketitle
%% The abstract (in this file, and that submitted as text to arXiv) should include the exact phrase
%% "fluid dynamics video" or "fluid dynamics videos"
\begin{abstract}
This is an entry for the Gallery of Fluid Motion of the 65st
Annual Meeting of the APS-DFD ( fluid dynamics video ). This
video shows the motion of levitated liquid droplets. The levitation is produced by the vertical vibration of a liquid container. We made visualizations of the motion of many droplets to study the formation of clusters and their stability. 
\end{abstract}
% main text

\section{Introduction}

If a liquid drop is deposited over a liquid surface,  the drop will, first, rebound, then arrest and eventually coalesce. Couder et al. \cite{couder} reported a technique to retard indefinitely the coalescence phase. By making the liquid container, the drop can be made to `sit' on top of the surface for a long time period. We built a similar experiment to study how several droplets cluster. 

\section{Experimental Conditions}
A short glass container was mounted on top of a commercial loudspeaker. The loudspeaker was fed with an amplified signal from a function generator. The frequency and amplitude of the signal were chosen such that large surface instabilities were not observed (Faraday waves). The liquid used was tap water. A small amount of liquid soap was used; we found that in this manner the drops where more stable. The process was filmed with a high speed camera.

\section{Videos}

Our video contributions can be found at:

\begin{itemize}
    \item \href{http://somewhere.net}{Video 1, mpeg4, full
resolution}

    \item \href{http://somewhere.net}{Video 2, mpeg2, low
resolution}

\end{itemize}

\end{document}